# Continuous variable entanglement with orbital angular momentum multiplexing in coherently prepared media

Fan Meng, Hao Zhu, Xin-Yao Huang[†], Guo-Feng Zhang[*]

*School of Physics, Beihang University, Beijing 102206, China*

**Abstract**

Quantum entanglement constitutes a pivotal resource, serving as a fundamental cornerstone within the field of quantum information science. In recent years, the study of vortex light entanglement has garnered widespread attention due to its unique structure and inherent advantages; however, the majority of these investigations are primarily focused on discrete variable (DV) systems. In this paper, we present a theoretical framework for generating vortex optical entanglement in coherently prepared media, employing continuous variable (CV) analysis and leveraging Raman scattering as an alternative to the conventional spontaneous parametric down-conversion (SPDC) method. The entanglement arises from the quantum correlation between the two light fields, induced by atomic coherence. Using numerical simulations, we thoroughly explore the impact of various tunable system parameters on the degree of entanglement, ultimately identifying the optimal conditions for maximal entanglement. Our findings offer a reference framework for vortex light entanglement, with potential implications across quantum teleportation, quantum key distribution, quantum computing, high-dimensional quantum information, and other related fields.

## 1. Introduction

Quantum entanglement constitutes a fundamental resource in quantum information science. Since Einstein, Podolsky, and Rosen introduced the concept of quantum entanglement in 1935 [1], it has persistently captivated the attention of the scientific community. The pioneering investigations into quantum entanglement were conducted within discrete variable (DV) quantum systems [2], which have subsequently spurred extensive theoretical and experimental research [3-8]. However, the probabilistic nature of DV systems poses a significant challenge, as the success rate diminishes with increasing entanglement scale, thereby impeding the realization of large-scale quantum entanglement. In contrast, continuous variable (CV) quantum systems [9] have emerged as natural platforms for realizing large-scale quantum entanglement, owing to their strengths in deterministic state preparation and quantum state conversion. In recent years, CV systems have sparked widespread interest, as evidenced by a plethora of studies. The experimental pursuit of CV quantum entanglement began with the generation of squeezed light [10-12], which serves as a foundational quantum resource. Subsequent research on CV entanglement has opened up a wealth of exciting possibilities [13-19].

Currently, entanglement can be realized across various quantum systems, including photon entanglement [20], atomic entanglement [21], ion entanglement [22], etc. Photons, characterized by rapid transmission, high bandwidth, and strong resistance to interference, are deemed exceptional information carriers, playing a crucial role in quantum communication and information processing. Vortex light, characterized by a unique degree of freedom, possesses a spiral phase factor, denoted as $\exp(il\varphi)$. Each photon is capable of carrying an orbital angular momentum (OAM) of $l\hbar$, where $l$ represents the topological charge, which can be either an integer or a non-integer. Consequently, vortex light emerges as an ideal information carrier, inherently suited for constructing high-dimensional quantum networks, thereby substantially enhancing both transmission capacity and information security [23-26]. The investigation of the entanglement properties of vortex light is significance. Anton Zeilinger et al. first realized the entanglement of two-photon OAMs via SPDC [27], catalyzing extensive research on photon OAM entanglement. The entanglement of photon OAM states was achieved in a hot atom ensemble through spontaneous four-wave mixing (SFWM) [28]. Quantum storage of three-dimensional entangled photon OAM was successfully demonstrated in rare-earth ion-doped crystals [29]. Experimental demonstration of multipartite quantum entanglement involving four-photon OAM was achieved [30]. Multi-degree-of-freedom super-entanglement and hybrid entanglement, incorporating both path and OAM, were experimentally established between two spatially separated atomic ensembles using quantum storage [31]. Through the use of a spatially non-degenerate OPO, the orthogonal entanglement between the first-order CV OAM states of two Laguerre-Gaussian modes was experimentally explored for the first time [32]. Optical OAM multiplexing and multi-part CV entanglement were achieved through a FWM process in hot $^{85}Rb$ vapor [33,34]. Other notable examples include [35-37].

---

[†] Corresponding author. E-mail: xinyaohuang@buaa.edu.cn
[*] Corresponding author. E-mail: gf1978zhang@buaa.edu.cn

From the aforementioned studies, it is evident that due to the inherently discrete nature of the OAM of vortex light, its entanglement can be explored from the DV perspective. Simultaneously, as an optical mode, vortex light exhibits characteristics of CV, allowing its entanglement to be investigated from the CV perspective as well. While numerous studies have explored the DV entanglement of vortex light, comparatively fewer investigations have addressed CV entanglement. Furthermore, the majority of the reported studies on vortex light entanglement are experimental in nature, with theoretical discussions remaining relatively scarce. This discrepancy can likely be attributed to the more established use of spatial light modulators (SLMs) as spatial post-selection tools [38] for generating vortex light and facilitating entanglement between the modes. In this work, we adopt the perspective of CV and propose a theoretical framework for generating vortex light entanglement within a three-level $\Lambda$-type system. In this framework, vortex light is employed instead of Gaussian light, enabling direct interaction with the medium without relying on spatial post-selection schemes. This methodology may offer advantages in enhancing the signal efficiency of vortex light while preserving mode purity. Specifically, we initially prepare the medium in its ground state and subsequently transition it to a coherent superposition state, followed by interaction with the vortex light. Under appropriately tunable system parameters, two entangled vortex lights are generated, and we identify the optimal entanglement configuration. The physical mechanism underlying this entanglement differs from traditional SPDC techniques [3,4]. This mechanism is rooted in the electromagnetically induced transparency (EIT) Raman scattering process [39]. Related studies have demonstrated that, compared to standard SPDC, the correlated photon source generated via EIT offers advantages such as narrow spectrum, high conversion efficiency, and extended coherence time and length [5,6,7]. Additionally, the incorporation of vortex light introduces a new dimension, providing a distinct advantage over conventional light. Our method for generating vortex light entanglement offers a paradigm for studying vortex light entanglement in CV, with potential applications in quantum information processing and quantum communication.

The structure of this paper is organized as follows: Sec. 2 presents a comprehensive theoretical derivation of the proposed system model. Sec. 3 details the numerical simulation results concerning the entanglement of vortex light, grounded in the theoretical framework. In Sec. 4, we provide an in-depth analysis of the physical mechanisms underpinning these numerical findings, accompanied by an approximate analytical expression of the entanglement. Finally, we conclude by summarizing our contributions and appending the detailed theoretical derivation process.

## 2. Theoretical model

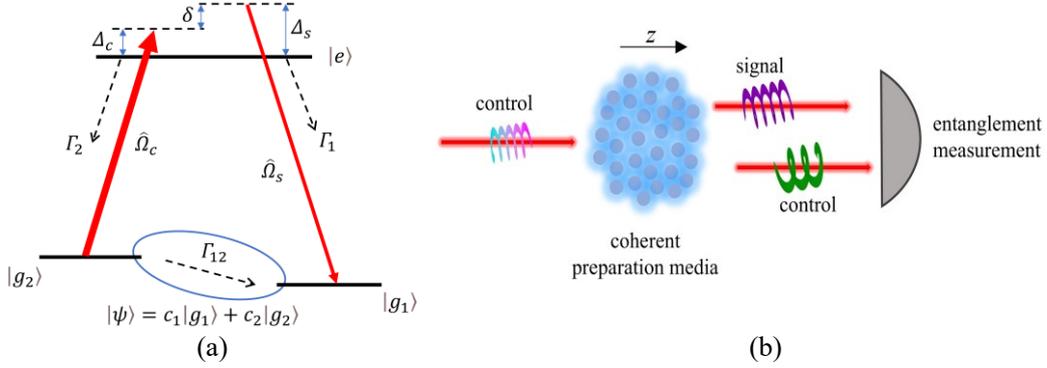

**Fig. 1.** (a) Schematic of the atomic three-level $\Lambda$-type system configuration. The two ground states of the atom are initially prepared in a coherent superposition state, with the strong control field $\Omega_c$ mediates the transition between energy levels $|g_2\rangle$ and $|e\rangle$, thereby generating a signal field $\Omega_s$ spanning from $|e\rangle$ to $|g_1\rangle$ through Raman scattering. The parameters $\Gamma_1$ and $\Gamma_2$ denote the decay rates of the upper energy levels; $\Delta_c$ and $\Delta_s$ refer to the single-photon detuning values, while $\delta = \Delta_s - \Delta_c$ specifies the binary detuning. Additionally, $\Gamma_{12}$ represents the ground state relaxation rate. (b) Interaction between the control field and the coherently prepared atomic ensemble, with entanglement measurements of the control and generated signal fields conducted at the output.

We consider an atomic ensemble configured in a three-level $\Lambda$-type scheme, as depicted in Fig. 1. The energy levels $|g_1\rangle$ and $|g_2\rangle$ represent the two ground states, while $|e\rangle$ denotes the excited state. $\Delta_c = \omega_c - \omega_{eg_2}$ and $\Delta_s = \omega_s - \omega_{eg_1}$ represent single-photon detuning, with $\omega_c$ and $\omega_s$ corresponding to the frequencies of the two light fields, and $\omega_{eg_1}$ and $\omega_{eg_2}$ denoting the transition frequencies between the corresponding atomic energy levels. The twin detuning, denoted as $\delta = \Delta_s - \Delta_c$, is a crucial parameter in the three-level $\Lambda$ system. The decay rates from the excited state $|e\rangle$ to the two ground states $|g_1\rangle$ and $|g_2\rangle$ are $\Gamma_1$ and $\Gamma_2$, respectively, assuming $\Gamma_1 = \Gamma_2 = \Gamma/2$, where $\Gamma$ represents the total decay rate of the upper energy level $|e\rangle$. $\Gamma_{12}$ denotes the decay rate of the particle number in the ground state. Additionally, the decoherence rate $\gamma_\phi$ between the two ground states, induced by atomic collisions and other factors, is also considered. Given the similarity in the functions of $\Gamma_{12}$ and $\gamma_\phi$, we simplify by defining $\gamma_\phi = 0$ and referring to $\Gamma_{12}$ as the relaxation rate [40]. We investigate the propagation of two laser fields with Rabi frequencies $\Omega_c$ and $\Omega_s$ within the medium, where $\Omega_s$ couples the energy levels $|g_1\rangle$ and $|e\rangle$, and $\Omega_c$ couples the levels $|g_2\rangle$ and $|e\rangle$. The Hamiltonian of such a system can be expressed as

$$\hat{H} = -\hbar\Delta_s\hat{\sigma}_{ee}(z,t) - \hbar\delta\hat{\sigma}_{g_2g_2}(z,t) - \frac{\hbar}{2}\left[\hat{\Omega}_s(z,t)\hat{\sigma}_{eg_1}(z,t) + \hat{\Omega}_c(z,t)\hat{\sigma}_{eg_2}(z,t) + \text{H.c}\right], \quad (1)$$

where $\hat{\Omega}_s(z,t) = g_s\hat{E}_s(z,t)$ and $\hat{\Omega}_c(z,t) = g_c\hat{E}_c(z,t)$, $g_s$ and $g_c$ are single-photon Rabi frequencies, assuming $g_s = g_c = g$, $\hat{E}_s$ and $\hat{E}_c$ are dimensionless light field operators that adhere to the bosonic commutation relation $\left[\hat{E}_\mu, \hat{E}_\mu^\dagger\right] = 1$, $\mu \in c, s$.

The dynamic evolution of atomic systems is governed by the Heisenberg-Langevin equation [19,41,42], expressed as follows

$$\frac{\partial}{\partial t}\hat{\sigma}_{\mu\nu} = \mathcal{L}_\sigma + \frac{1}{i\hbar}\left[\hat{\sigma}_{\mu\nu}, \hat{H}\right] + \hat{F}_{\mu\nu} \quad (\mu,\nu \in e, g_1, g_2). \tag{2}$$

The first term on the right-hand side of the equation represents the damping component, the second term captures the primary dynamic process, and the final term corresponds to the Langevin noise operator, which adheres to the fluctuation-dissipation theorem. Introducing this operator is essential to preserve the commutation relations among the operators.

The propagation of light fields through a medium is described by the Maxwell-Schrödinger equation [19,41,42] as follows

$$\left(\frac{\partial}{\partial z} + c^{-1}\frac{\partial}{\partial t}\right)\hat{\Omega}_s = i\frac{\Gamma\alpha}{2L}\hat{\sigma}_{g_1 e}, \tag{3}$$

$$\left(\frac{\partial}{\partial z} + c^{-1}\frac{\partial}{\partial t}\right)\hat{\Omega}_c = i\frac{\Gamma\alpha}{2L}\hat{\sigma}_{g_2 e}, \tag{4}$$

where $\alpha = 4g^2 NL/c\Gamma$ represents the optical density of the medium, $N$ denotes the total number of atoms in the ensemble, and $L$ refers to the length of the medium.

Solving the Eqs. (1)-(4) yields a set of coupling equations that describe the interactions between atomic operators and field operators. To solve this set of coupled equations, we employ the mean-field approximation, decomposing each operator $\hat{O}$ into a mean-field component $O$ and a quantum fluctuation component $\hat{o}$, i.e., $\hat{O} = O + \hat{o}$. Consequently, for atomic operators, we denote $\hat{\sigma}_{\mu\nu} = \sigma_{\mu\nu} + \hat{s}_{\mu\nu}$ $(\mu,\nu \in e, g_1, g_2)$, and for light field operators $\hat{E}_\mu = E_\mu + \hat{a}_\mu$ $(\mu \in c,s)$, where $\sigma_{\mu\nu}$ and $E_\mu$ represent the mean-field components of atoms and light fields, respectively, while $\hat{s}_{\mu\nu}$ and $\hat{a}_\mu$ represent the quantum fluctuation components of atoms and light fields, respectively. Regarding open quantum dynamics, the accuracy of the mean-field approach has been rigorously demonstrated in systems characterized by collective jump operators and Hamiltonians that exhibit all-to-all interactions among subsystems, as well as in various formulations of spin-boson models [43]. The validity of the mean-field approach in specific open quantum systems has also been investigated through numerical simulations [44,45]. To clearly elucidate the problem under investigation, we will separately solve the mean-field component and the quantum fluctuation component in Sec. 2. Detailed specifics of the solution process are provided in the appendix.

## 2.1 Theoretical model for the generation of two vortex beams in a coherently prepared medium

Under the mean-field approximation, the mean-field components of the atomic operators $\hat{\sigma}_{g_1 e}$ and $\hat{\sigma}_{g_2 e}$ can be formulated as follows

$$\frac{\partial}{\partial t}\sigma_{g_1 e} = \left(-\frac{\Gamma}{2} + i\Delta_s\right)\sigma_{g_1 e} + i\frac{\Omega_s}{2}\left(\sigma_{g_1 g_1} - \sigma_{ee}\right) + i\frac{\Omega_c}{2}\sigma_{g_1 g_2}, \tag{5}$$

$$\frac{\partial}{\partial t}\sigma_{g_2 e} = \left(-\frac{\Gamma + \Gamma_{12}}{2} + i\Delta_c\right)\sigma_{g_2 e} + i\frac{\Omega_c}{2}\left(\sigma_{g_2 g_2} - \sigma_{ee}\right) + i\frac{\Omega_s}{2}\sigma_{g_2 g_1}. \tag{6}$$

The Maxwell-Schrödinger equation, governing the propagation of light within a medium, can be expressed in terms of its mean-field components as

$$\left(\frac{\partial}{\partial z}+c^{-1}\frac{\partial}{\partial t}\right)\Omega_s = i\frac{\Gamma\alpha}{2L}\sigma_{g_1 e}, \tag{7}$$

$$\left(\frac{\partial}{\partial z}+c^{-1}\frac{\partial}{\partial t}\right)\Omega_c = i\frac{\Gamma\alpha}{2L}\sigma_{g_2 e}. \tag{8}$$

Consider an atom initially prepared in a coherent superposition of two ground states:

$$|\psi\rangle = c_1|g_1\rangle + c_2|g_2\rangle. \tag{9}$$

By focusing on a weak interaction between fields and atoms, we approximate $\sigma_{ee} \approx 0$, $\sigma_{g_1 g_1} \approx |c_1|^2$, $\sigma_{g_2 g_2} \approx |c_2|^2$, $\sigma_{g_1 g_2} \approx c_1^* c_2$. However, for sufficiently long media, the transmission of light fields can reach a stable state, at which point the assumption of weak light-atom interaction becomes less restrictive [46]. We focus on the Raman process, where initially only the light field $\Omega_c$ is present at the medium entrance $z=0$, with the conditions $\Omega_c(0)=\Omega$ and $\Omega_s(0)=0$. Under the given conditions, the steady-state solution of Eqs. (5)-(9) is derived, with the time-derivative terms neglected, yielding the following expression for the propagation of the light field in the medium:

$$\Omega_c(z) = \frac{\Omega}{W}\left(\beta_2|c_2|^2 e^{-iWz} + \beta_1|c_1|^2\right), \tag{10}$$

$$\Omega_s(z) = \frac{\Omega}{W}c_1^* c_2 \beta_1 \left(e^{-iWz} - 1\right), \tag{11}$$

where $\beta_1 = \frac{\Gamma\alpha}{2L(2\Delta_s + i\Gamma)}$, $\beta_2 = \frac{\Gamma\alpha}{2L[2\Delta_c + i(\Gamma + \Gamma_{12})]}$, and $W = \beta_1|c_1|^2 + \beta_2|c_2|^2$.

Letting the initial incident beam $\Omega_c$ be vortex light, that is, $\Omega_c(0) = \Omega = |\Omega|e^{il\varphi}$, where $l$ denotes the topological charge of the vortex and $\varphi$ is the azimuthal angle around the z-axis. Considering the most common Laguerre-Gaussian vortex form, its transverse profile can be expressed as

$$|\Omega| = \varepsilon\left(\frac{r}{w}\right)^{|l|} e^{-r^2/w^2}, \tag{12}$$

where $\varepsilon$ denotes the amplitude of the vortex light, $r$ represents the cylinder radius, and $w$ indicates the beam waist. Analysis of Eqs. (10) and (11) reveals that the generated light field $\Omega_s(z) \sim |\Omega|e^{il\varphi}$ demonstrates that by initially preparing the medium's two ground states in a coherent superposition state, we can utilize a vortex beam to interact with the medium, thereby generating two vortex beams at the medium's exit. It is important to note that this process occurs only when the two low-energy states of the atom are initially in a coherent superposition; otherwise, phase fluctuations will average out the generation process, as illustrated on the right-hand side of Eq. (11). From Eqs. (10) and (11), it is evident that as the two beams propagate through the medium, losses are encountered; however, these losses are confined to the medium's entrance. Once the beams have traversed a certain distance, these losses dissipate, due to the establishment of EIT. For simplicity, we consider the special scenario where $\delta = 0$ and $\Gamma_{12} = 0$, in which $\beta_1 = \beta_2 = W = \alpha/2iL_{abs}$ (with $L_{abs} = L/\alpha$ representing the absorption length). If the medium's optical density is sufficiently large, i.e., $\alpha \gg 1$, the effective absorption length $L_{abs} \ll L$ occupies only a small fraction of the entire medium. For $z \gg L_{abs}$, we present the propagation dynamics of the two fields following the establishment of EIT:

$$\Omega_c(z \gg L_{abs}) = \Omega|c_1|^2 = \varepsilon\left(\frac{r}{w}\right)^{|l|} e^{-r^2/w^2}|c_1|^2 e^{il\varphi}, \tag{13}$$

$$\Omega_s(z \gg L_{abs}) = -\Omega c_1^* c_2 = -\varepsilon \left(\frac{r}{w}\right)^{|l|} e^{-r^2/w^2} c_1^* c_2 e^{il\varphi}. \quad (14)$$

While losses initially manifest at the medium's entrance, as the beams penetrate deeper, and atoms transition into the dark state, these losses gradually abate [47]. It is important to note that the above results are obtained under ideal EIT conditions, where the detuning parameters $\Delta_c = \Delta_s = 0$ and the ground-state relaxation rate $\Gamma_{12} = 0$. Given that the entanglement investigated in this study necessitates tunable system parameters, the conditions described do not strictly adhere to standard EIT. However, within the tuning range of entanglement, the EIT condition can still be effectively maintained. Fig. 2 provides a clear depiction of the propagation of the light field within the medium. In this figure, we present the variation in the incident field intensity $|\Omega_c(z)|^2/|\Omega_c(0)|^2$ and the generated field intensity $|\Omega_s(z)|^2/|\Omega_c(0)|^2$ as a function of the dimensionless length $z/L$ under the standard EIT conditions: $\delta = 0$, $\Gamma_{12} = 0$, $\alpha = 50$, and $c_1 = c_2 = 1/\sqrt{2}$ in panel 2(a), and the entanglement tuning range: $\delta = 0.5$, $\Gamma_{12} = 0.5$, $\alpha = 50$, and $c_1 = c_2 = 1/\sqrt{2}$ in panel 2(b).

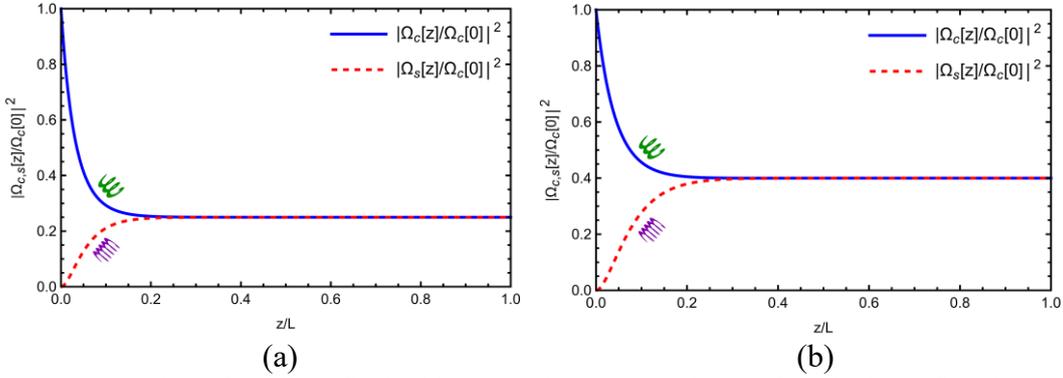

(a)                                   (b)

**Fig. 2.** Propagation of the light field within a medium. Dependence of the incident field intensity $|\Omega_c(z)|^2/|\Omega_c(0)|^2$ and generated field intensity $|\Omega_s(z)|^2/|\Omega_c(0)|^2$ on the dimensionless length $z/L$ for (a) the standard EIT conditions: $\delta = 0$, $\Gamma_{12} = 0$, $\alpha = 50$, $c_1 = c_2 = 1/\sqrt{2}$, and (b) the entanglement tuning range: $\delta = 0.5$, $\Gamma_{12} = 0.5$, $\alpha = 50$, $c_1 = c_2 = 1/\sqrt{2}$.

## 2.2 Theoretical model for the entanglement of two vortex beams

Subsequently, we aim to investigate the potential entanglement of the two vortex beams generated by this method, contingent upon appropriate system parameters. Indeed, our findings affirm that entanglement can be achieved. The proposed physical mechanism for generating vortex optical entanglement is based on the Raman scattering process, distinguishing it from conventional SPDC methods.

Given that light field entanglement is inherently a quantum phenomenon, it is imperative to account for the quantum fluctuation components within the equation. To this end, the mean-field approximation is employed. By neglecting the higher-order terms, we derive the kinematic equation of the atomic fluctuation operator as follows

$$\frac{\partial}{\partial t}\hat{s}_{eg_1} = \left(-\frac{\Gamma}{2} - i\Delta_s\right)\hat{s}_{eg_1} - i\frac{\Omega_s^*}{2}(\hat{s}_{g_1g_1} - \hat{s}_{ee}) - i\frac{g}{2}(\sigma_{g_1g_1} - \sigma_{ee})\hat{a}_s^+ - i\frac{\Omega_c^*}{2}\hat{s}_{g_2g_1} - i\frac{g}{2}\sigma_{g_2g_1}\hat{a}_c^+ + \hat{F}_{eg_1}, \quad (15)$$

$$\frac{\partial}{\partial t}\hat{s}_{eg_2} = \left(-\frac{\Gamma + \Gamma_{12}}{2} - i\Delta_c\right)\hat{s}_{eg_2} - i\frac{\Omega_c^*}{2}(\hat{s}_{g_2g_2} - \hat{s}_{ee}) - i\frac{g}{2}(\sigma_{g_2g_2} - \sigma_{ee})\hat{a}_c^+ - i\frac{\Omega_s^*}{2}\hat{s}_{g_1g_2} - i\frac{g}{2}\sigma_{g_1g_2}\hat{a}_s^+ + \hat{F}_{eg_2}, \quad (16)$$

$$\frac{\partial}{\partial t}\hat{s}_{g_2g_1} = \left(-\frac{\Gamma_{12}}{2} - \gamma_\phi - i\delta\right)\hat{s}_{g_2g_1} + i\frac{\Omega_s^*}{2}\hat{s}_{g_2e} + i\frac{g}{2}\sigma_{g_2e}\hat{a}_s^+ - i\frac{g}{2}\sigma_{eg_1}\hat{a}_c - \frac{i}{2}\Omega_c\hat{s}_{eg_1} + \hat{F}_{g_2g_1}, \quad (17)$$

$$\frac{\partial}{\partial t}\hat{s}_{g_1g_1} = \Gamma_1\hat{s}_{ee} + \Gamma_{12}\hat{s}_{g_2g_2} - i\frac{\Omega_s}{2}\hat{s}_{eg_1} - i\frac{g}{2}\sigma_{eg_1}\hat{a}_s + i\frac{\Omega_s^*}{2}\hat{s}_{g_1e} + i\frac{g}{2}\sigma_{g_1e}\hat{a}_s^+ + \hat{F}_{g_1g_1}, \quad (18)$$

$$\frac{\partial}{\partial t}\hat{s}_{g_2g_2} = \Gamma_2\hat{s}_{ee} - \Gamma_{12}\hat{s}_{g_2g_2} - i\frac{\Omega_c}{2}\hat{s}_{eg_2} - i\frac{g}{2}\sigma_{eg_2}\hat{a}_c + i\frac{\Omega_c^*}{2}\hat{s}_{g_2e} + i\frac{g}{2}\sigma_{g_2e}\hat{a}_c^+ + \hat{F}_{g_2g_2}, \quad (19)$$

$$\frac{\partial}{\partial t}\hat{s}_{ee} = -\Gamma\hat{s}_{ee} + i\frac{\Omega_s}{2}\hat{s}_{eg_1} + i\frac{g}{2}\sigma_{eg_1}\hat{a}_s + i\frac{\Omega_c}{2}\hat{s}_{eg_2} + i\frac{g}{2}\sigma_{eg_2}\hat{a}_c - i\frac{\Omega_s^*}{2}\hat{s}_{g_1e} - i\frac{g}{2}\sigma_{g_1e}\hat{a}_s^+ - i\frac{\Omega_c^*}{2}\hat{s}_{g_2e} - i\frac{g}{2}\sigma_{g_2e}\hat{a}_c^+ + \hat{F}_{ee}, \quad (20)$$

$$\frac{\partial}{\partial t}\hat{s}_{g_1g_2} = \frac{\partial}{\partial t}\hat{s}_{g_2g_1}^\dagger, \quad (21)$$

$$\frac{\partial}{\partial t}\hat{s}_{g_2e} = \frac{\partial}{\partial t}\hat{s}_{eg_2}^\dagger, \quad (22)$$

$$\frac{\partial}{\partial t}\hat{s}_{g_1e} = \frac{\partial}{\partial t}\hat{s}_{eg_1}^\dagger. \quad (23)$$

The quantum fluctuation component of the Maxwell-Schrödinger equations governing the propagation of light field in a medium is expressed as

$$\left(\frac{\partial}{\partial z} + c^{-1}\frac{\partial}{\partial t}\right)g\hat{a}_s = i\frac{\Gamma\alpha}{2L}\hat{s}_{g_1e}, \quad (24)$$

$$\left(\frac{\partial}{\partial z} + c^{-1}\frac{\partial}{\partial t}\right)g\hat{a}_c = i\frac{\Gamma\alpha}{2L}\hat{s}_{g_2e}. \quad (25)$$

Substituting Eqs. (15)-(23) into Eqs. (24) and (25) yields a set of coupled equations involving the light field fluctuation operators $\hat{a}_s$ and $\hat{a}_c$. Solving this set of coupled equations allows us to investigate the properties of light field entanglement. The underlying physical mechanisms that lead to the generation of continuous variable entanglement in the system under study will be elaborated upon in detail in the discussion section.

In the context of quantum entanglement quantification, various entanglement criteria have been developed based on Bell's inequality to assess entanglement. In CV quantum systems, two widely recognized standards are commonly employed [48, 49]. Here we utilize the Duan-Giedke-Cirac-Zoller (DGCZ) entanglement criterion. According to the DGCZ criterion, the two light fields $\Omega_s$ and $\Omega_c$ must satisfy the following condition:

$$V(\theta) = \Delta^2\left[\hat{X}_c(\theta) + \hat{X}_s(\theta)\right] + \Delta^2\left[\hat{Y}_c(\theta) - \hat{Y}_s(\theta)\right] < 4, \quad (26)$$

where $\hat{X}_\mu(\theta) = \hat{a}_\mu e^{-i\theta} + \hat{a}_\mu^\dagger e^{i\theta}$ and $\hat{Y}_\mu(\theta) = -i(\hat{a}_\mu e^{-i\theta} - \hat{a}_\mu^\dagger e^{i\theta})$, $(\mu \in c, s)$ represent two orthogonal operators of the light field, and $\theta$ denotes the orthogonal angle. The two physical quantities $\hat{X}$ and $\hat{Y}$, which signify entanglement in the optical mode, are analogous to the position and momentum of particles, representing two continuous quantum variables [50]. Consequently, our system is classified within the CV domain. It is worth noting that previous studies have explored vortex light continuous variable entanglement through the application of the DGCZ criterion [32, 51].

By selecting the optimal orthogonal angle $\theta_{opt}$, we derive an expression that minimizes the entanglement degree $V$ as expressed by the light field operator:

$$V = 4\left[1 + \langle\hat{a}_c^\dagger\hat{a}_c\rangle + \langle\hat{a}_s^\dagger\hat{a}_s\rangle - 2|\langle\hat{a}_s\hat{a}_c\rangle|\right]. \quad (27)$$

We consider the scenario of stable light field transmission within the medium. By solving the time-independent steady-state equations derived from Eqs. (24) and (25) and substituting them into Eq. (27), we examine the entanglement properties of the system. The degree of entanglement is contingent upon certain adjustable parameters. In Sec. 3, we will present the entanglement results for varying physical parameters and compare the corresponding entanglement degrees.

## 3. Numerical results

Based on the theoretical calculations outlined in Sec. 2, the entanglement of vortex beams is determined by the two-photon detuning $\delta$, ground state relaxation rate $\Gamma_{12}$, initial input field Rabi frequency $\Omega_c$, and optical density $\alpha$. For simplicity, we assume $\Delta_s = -\Delta_c = \delta/2$. In this section, we analyze how these parameters influence entanglement. To delve deeper into this problem, we will examine the influence of these parameters from three perspectives: **3.1.** Considering only two-photon detuning with the ground state relaxation rate set to zero; **3.2.** Considering only the ground state relaxation rate with the two-photon detuning set to zero; **3.3.** Simultaneously considering both two-photon detuning and ground state relaxation rate. Through this investigation, we identified the optimal entanglement within the tunable parameter range.

### 3.1. Two-photon detuning scheme

In the two-photon detuning scenario, to ensure that the entanglement arises solely from the two-photon detuning $\delta$ and is unaffected by the ground state relaxation rate, we set $\Gamma_{12} = 0$. Fig. 3(a) illustrates the relationship between the entanglement measure $V$ and the two-photon detuning $\delta$, with the optical density set to $\alpha = 50$, $\alpha = 100$, and $\alpha = 200$, respectively. For given values of the input field Rabi frequency $\Omega_c$ and optical density $\alpha$, we can identify an optimal two-photon detuning $\delta_{opt}$ that maximizes the entanglement of the two output vortex beams ($V$ is minimized, i.e., $V_{opt,\delta}$). With fixed values of optical density $\alpha$ and two-photon detuning $\delta$, there exists an optimal Rabi frequency $\Omega_{c,opt}$ of the initial input field, which makes the output entanglement optimal $V_{opt,\Omega_c}$, as shown in Fig. 3 (b).

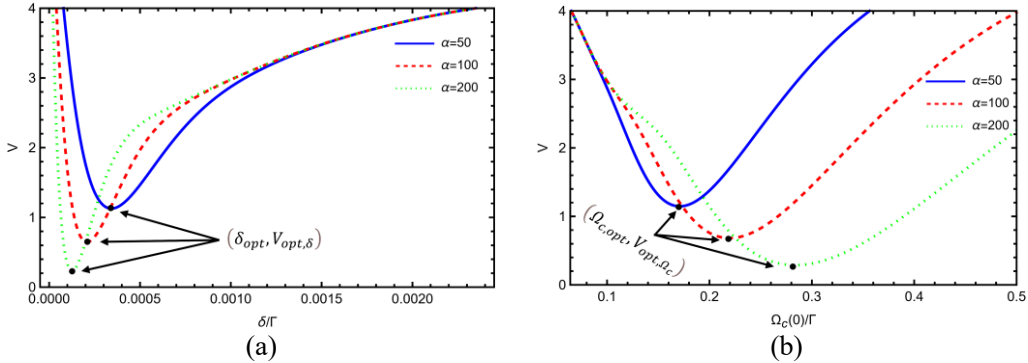

(a)            (b)

**Fig. 3.** (a) Dependence of the entanglement on the two-photon detuning $\delta$ for $\alpha = 50, 100, 200$ and $\Omega_c = 0.1\Gamma$. (b) Dependence of the entanglement on the initial input field Rabi frequency $\Omega_c$ for $\alpha = 50, 100, 200$ and $\delta = 0.001\Gamma$. In both figures, the ground state relaxation rate is fixed at $\Gamma_{12} = 0$.

From the above results, it is evident that within the tunable range, the entanglement $V$ exhibits significant dependence on both the two-photon detuning $\delta$ and the input field Rabi frequency $\Omega_c$, with an optimal value minimizing the entanglement $V$. Therefore, considering a fixed optical density $\alpha$, we can determine the dependence of entanglement measure $V$ on the two-photon detuning $\delta$ and the input field Rabi frequency $\Omega_c$, allowing us to identify the optimal entanglement value $V_{opt,(\delta,\Omega_c)}$, as shown in Fig. 4. It is important to note that we focus on the case of low optical density, specifically $\alpha = 50$; further increases in the optical density of the medium enhance the entanglement of the system. In Fig. 4, to ensure that the entanglement originates from the two-photon detuning $\delta$, we assume $\Gamma_{12} = 0$.

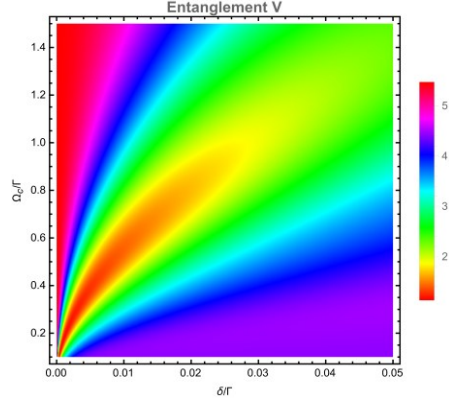

**Fig. 4.** Dependence of the entanglement on the two-photon detuning $\delta$ and input field Rabi frequency $\Omega_c$ for $\alpha = 50$. In this figure, the ground state relaxation rate is fixed at $\Gamma_{12} = 0$.

### 3.2. Ground state relaxation rate scheme

In the ground state relaxation rate scheme, to ensure that the entanglement solely arises from the ground state relaxation rate $\Gamma_{12}$, we set $\delta = 0$. Fig. 5 (a) illustrates the relationship between the entanglement measure $V$ and the ground state relaxation rate $\Gamma_{12}$, with the optical density set to $\alpha = 50$, $\alpha = 100$, and $\alpha = 200$, respectively. For fixed values of the input field Rabi frequency $\Omega_c$ and the optical density $\alpha$, by scanning the ground state relaxation rate $\Gamma_{12}$, we can identify an optimal $\Gamma_{12,opt}$ that minimizes the output entanglement $V$, i.e., $V_{opt,\Gamma_{12}}$. With the optical density $\alpha$ and ground state relaxation rate $\Gamma_{12}$ fixed, there exists an optimal Rabi frequency $\Omega_{c,opt}$ for the initial input field that minimizes the output entanglement $V_{opt,\Omega_c}$, as depicted in Fig. 5 (b).

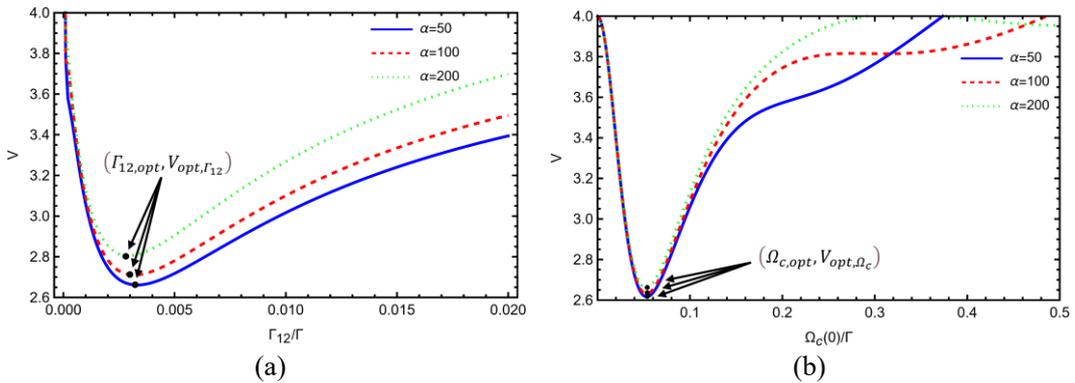

(a)            (b)

**Fig. 5.** (a) Dependence of the entanglement on the ground state relaxation rate $\Gamma_{12}$ for $\alpha = 50,100,200$ and $\Omega_c = 0.1\Gamma$. (b) Dependence of the entanglement on the input field Rabi frequency $\Omega_c$ for $\alpha = 50,100,200$ and $\Gamma_{12} = 0.001\Gamma$. In both figures, the two-photon detuning is fixed at $\delta = 0$.

Similarly, the results indicate that the entanglement $V$ exhibits significant variation with both the ground state relaxation rate $\Gamma_{12}$ and the input field Rabi frequency $\Omega_c$, with an optimal value minimizing the entanglement $V$. This consideration leads us to explore the dependence of the entanglement measure $V$ on the ground state relaxation rate $\Gamma_{12}$ and input field Rabi frequency $\Omega_c$. Given a fixed optical density $\alpha$, we can determine the optimal $\Gamma_{12,opt}$ and $\Omega_{c,opt}$ that minimize the entanglement $V$, i.e., $V_{opt,(\Gamma_{12},\Omega_c)}$, as illustrated in Fig. 6. Note that we have considered only the case of low optical density, specifically $\alpha = 50$. In Fig. 6, to ensure that the entanglement originates from the ground state relaxation rate $\Gamma_{12}$, we assume $\delta = 0$.

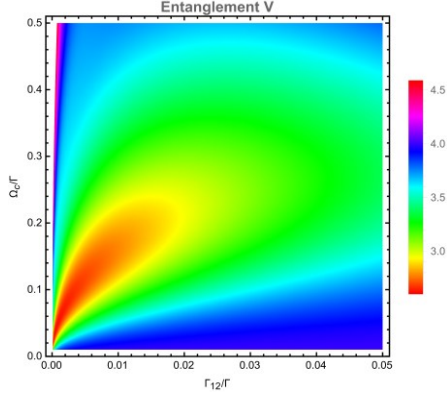

**Fig. 6.** Dependence of the entanglement on the ground state relaxation rate $\Gamma_{12}$ and input field Rabi frequency $\Omega_c$ for $\alpha = 50$. The two-photon detuning is set to $\delta = 0$ in this figure.

Initially, we provide a brief discussion of the numerical results for the two entanglement schemes described above, with a more detailed analysis of their underlying physical mechanisms presented in Sec. 4. As shown in Fig. 3 (a) and 5 (a), when $\delta = \Gamma_{12} = 0$, the output field exhibits no entanglement; however, when $\delta$ and $\Gamma_{12}$ exist and assume appropriate values, entanglement occurs between the two output fields. Comparing the entanglement generated by the two schemes, we observe that the two-photon detuning scheme yields greater entanglement. As evident from Fig. 4 and 6, with constant optical density, the tunable range for optimal entanglement $V_{opt,(\delta,\Omega_c)}$ generated by the two-photon detuning scheme is broader, whereas the tunable range for optimal entanglement $V_{opt,(\Gamma_{12},\Omega_c)}$ generated by the ground state relaxation rate scheme is narrower. Besides being related to $\delta$, $\Gamma_{12}$, and $\Omega_c$, the degree of entanglement is also influenced by the optical density $\alpha$ of the medium. Fig. 3 and 5 reveal that the optimal entanglement $V_{opt}$ in the two-photon detuning scheme is sensitive to variations in optical density $\alpha$, with the entanglement $V$ decreasing as $\alpha$ increases. Therefore, in the two-photon detuning scheme, the optical density can enhance the output entanglement. In contrast, the optimal entanglement $V_{opt}$ in the ground state relaxation rate scheme shows minimal sensitivity to changes in optical density $\alpha$.

### 3.3. The scenario where two-photon detuning and ground-state relaxation rate coexist

From the preceding analysis, it is evident that two-photon detuning $\delta$ and ground state relaxation rate $\Gamma_{12}$ are critical parameters in generating entanglement, with both schemes allowing the determination of optimal entanglement within their respective tunable ranges. Let us now consider a more generalized scenario, wherein both two-photon detuning and ground state relaxation rate simultaneously influence entanglement, and seek to determine the optimal entanglement. Fig. 7 (a) illustrates the dependence of entanglement $V$ on the two-photon detuning $\delta$ and ground state relaxation rate $\Gamma_{12}$. For a fixed input field Rabi frequency $\Omega_c$ and optical density $\alpha$, scanning the two-photon detuning $\delta$ and ground state relaxation rate $\Gamma_{12}$ allows us to identify an optimal $(\delta_{opt}, \Gamma_{12,opt})$ that minimizes the entanglement $V$, i.e., $V_{opt,(\delta,\Gamma_{12})}$. Given the optical density $\alpha$, two-photon detuning $\delta$, and ground state relaxation rate $\Gamma_{12}$, an optimal Rabi frequency $\Omega_{c,opt}$ of the initial input field exists, which yields optimal output entanglement $V_{opt,\Omega_c}$, as depicted in Fig. 7 (b).

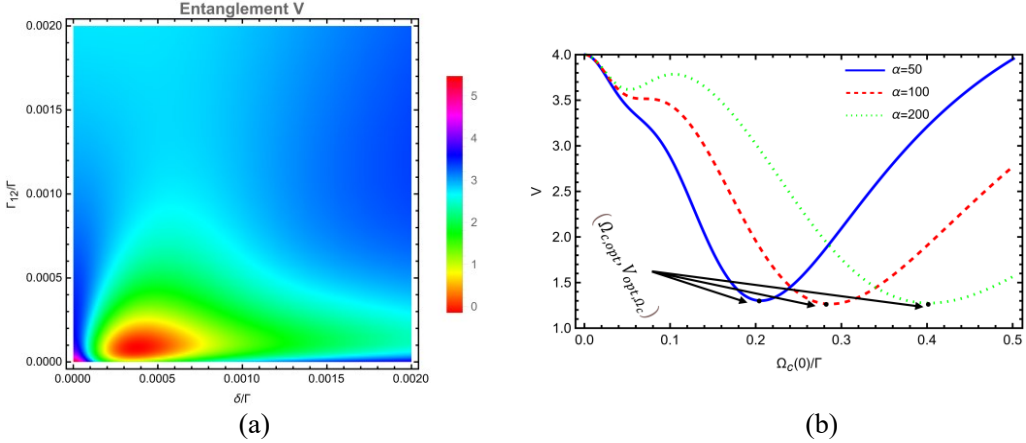

(a) (b)

**Fig. 7.** (a) Dependence of the entanglement on the two-photon detuning $\delta$ and ground state relaxation rate $\Gamma_{12}$ for $\alpha = 50$ and $\Omega_c = 0.1\Gamma$. (b) Dependence of the entanglement on the input field Rabi frequency $\Omega_c$ for $\alpha = 50, 100, 200$, $\delta = 0.001\Gamma$ and $\Gamma_{12} = 0.001\Gamma$.

Similarly, the preceding discussion reveals that entanglement $V$ varies significantly with two-photon detuning $\delta$, ground state relaxation rate $\Gamma_{12}$, and input field Rabi frequency $\Omega_c$, each having an optimal value that minimizes the entanglement $V$. This observation prompts an exploration of the dependence of entanglement $V$ on the two-photon detuning $\delta$, ground state relaxation rate $\Gamma_{12}$ and input field Rabi frequency $\Omega_c$. With a fixed optical density $\alpha$, we can determine the optimal $(\delta_{opt}, \Gamma_{12,opt}, \Omega_{c,opt})$ that minimizes the entanglement $V$, i.e., $V_{opt,(\delta,\Gamma_{12},\Omega_c)}$, as illustrated in Fig. 8. Panel (a) of Fig. 8 displays the optimal entanglement range corresponding to the relevant parameters, while panel (b) presents a cross-sectional view of this optimal entanglement region.

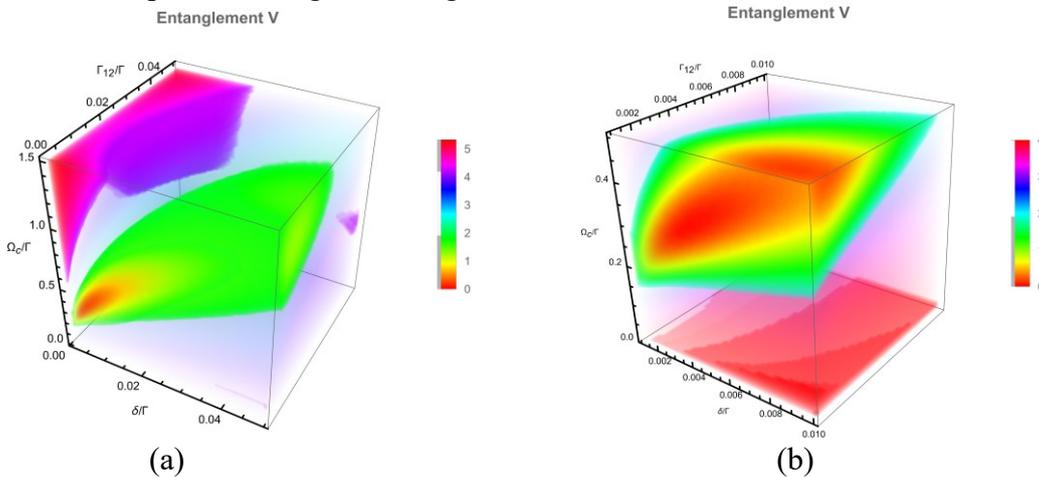

(a) (b)

**Fig. 8.** Dependence of the entanglement on the two-photon detuning $\delta$, ground state relaxation rate $\Gamma_{12}$ and input field Rabi frequency $\Omega_c$ for $\alpha = 50$. (a) The range of optimal entanglement. (b) The profile of optimal entanglement.

As shown in Fig. 7(a), no entanglement is observed between the output fields when $\delta = \Gamma_{12} = 0$, aligning with the outcomes in scenarios involving solely two-photon detuning or ground state relaxation. However, when $\delta$ and $\Gamma_{12}$ are within the appropriate tunable range, the optimal entanglement $V_{opt,(\delta,\Gamma_{12})}$ can be found, which can closely approach the ideal entanglement ($V = 0$). In contrast to the two-photon detuning scheme, Figure 7(b) illustrates that when the ground-state relaxation rate $\Gamma_{12}$ ($\Gamma_{12} = 0.001\Gamma$) is present, the variation of the optimal entanglement $V_{opt,\Omega_c}$ with optical density $\alpha$ is suppressed. Specifically, the entanglement $V$ becomes insensitive to increases in the optical density $\alpha$, which further supports the notion that entanglement varies slowly with optical density in the previously proposed ground-state relaxation rate scheme. Fig. 8(a) demonstrates that by adjusting the three parameters $\delta$, $\Gamma_{12}$, and $\Omega_c$, the optimal entanglement $V_{opt,(\delta,\Gamma_{12},\Omega_c)}$ can be located within a specific range, and the resulting entanglement can closely approximate the ideal entanglement. Notably, even at a relatively low optical density $\alpha = 50$, our scheme can achieve a high degree of entanglement. As previously discussed, increasing optical density results in a reduction of the entanglement $V$ overall. Therefore, the entanglement can be further optimized by increasing the optical density. Currently, an ultra-high optical density of $\alpha = 1000$ can be achieved experimentally [52,53].

## 4. Discussion

This paper primarily investigates the entanglement properties of two vortex beams within a coherently prepared medium. In Sec. 2, we present a theoretical framework and derive the dynamical behavior of atomic and light field operators. In Sec. 3, we provide numerical simulation results, elucidating the dependence of light field entanglement on various tunable system parameters derived from theoretical calculations. To gain a comprehensive understanding of the physical principles underlying the numerical results, we proceed with a detailed analysis of the system. From Eqs. (15) to (25), we obtain

$$\frac{\partial}{\partial \xi}\hat{a}_s = P_1\hat{a}_s + Q_1\hat{a}_s^\dagger + R_1\hat{a}_c + S_1\hat{a}_c^\dagger + \hat{n}_s, \tag{28}$$

$$\frac{\partial}{\partial \xi}\hat{a}_c = P_2\hat{a}_s + Q_2\hat{a}_s^\dagger + R_2\hat{a}_c + S_2\hat{a}_c^\dagger + \hat{n}_c, \tag{29}$$

where $\xi = z/L$ represents the dimensionless length, $P_i, Q_i, R_i, S_i \ (i=1,2)$ denote the corresponding coefficients, and $\hat{n}_s$ and $\hat{n}_c$ signify the Langevin noise operators. We now proceed to discuss in detail the physical significance of these coefficients. Firstly, the terms $P_1$ and $R_2$ correspond to the self-phase coupling process, a mechanism that does not generate entanglement. Secondly, the terms $Q_1$ and $S_2$ denote the single-mode compression process, corresponding to two independently squeezed light fields. The contribution of these terms to the entanglement is expressed as $V = 4\cosh(2|S_2|\xi)$, under the assumption that $S_2 = Q_1$ and remains constant with respect to $\xi$. Thirdly, the terms $R_1$ and $P_2$ signify the cross-phase coupling process, which does not result in entanglement at the output. Subsequently, the terms $S_1$ and $Q_2$ represent the dual-mode compression process, a key mechanism for generating entanglement. By considering only the $S_1$ and $Q_2$ terms, the output entanglement can be ideally represented as $V = 4e^{-2|S_1|\xi}$, under the assumption that $S_1 = Q_2$ and remains invariant with respect to $\xi$. Ultimately, the presence of Langevin noise terms, denoted as $\hat{n}_s$ and $\hat{n}_c$, leads to both the degradation and disruption of entanglement, thereby resulting in a reduction of its overall degree.

The optimal entanglement between light fields arises from the interplay between two-mode compression and other competing effects, including single-mode compression, self-phase coupling, and cross-phase coupling. Although these interactions do not directly generate entanglement, they compete with the two-mode compression process, ultimately leading to the establishment of a minimum point of entanglement $V$. Specifically, when the two-photon detuning $\delta$ and the ground state relaxation rate $\Gamma_{12}$ are zero, no entanglement exists between the control and signal fields. However, as $\delta$ and $\Gamma_{12}$ increase, the entanglement between the light fields is significantly enhanced, ultimately leading to near-ideal entanglement ($V = 0$). This phenomenon occurs because, when both $\delta$ and $\Gamma_{12}$ are zero, standard EIT is established, with $\sigma_{g_1 e}$ and $\sigma_{g_2 e}$ vanishing, causing the atoms to remain in the dark state. As a result, the medium becomes completely transparent to both the control and signal fields, and no entanglement is established between the two light fields. However, as $\delta$ and $\Gamma_{12}$ increase, the atoms deviate from the dark state, leading to mutual coupling between the two light fields and the generation of quantum entanglement. When $\delta$ and $\Gamma_{12}$ become sufficiently large, dissipation dominates, and the entanglement is weakened and eventually disappears. In the two-photon detuning scheme, the entanglement coefficient approximately follows the relationship $|S_1| \approx |Q_2| \approx [\alpha \Gamma \delta / (2\Omega^2)] \exp\{\Gamma^2 \alpha \xi / (\Gamma^2 + \delta^2)\}$, while for the ground state relaxation rate scheme, the entanglement coefficient similarly follows $|S_1| \approx |Q_2| \approx [\alpha \Gamma \Gamma_{12} / (4\Omega^2)] \exp\{\alpha (2\Gamma + \Gamma_{12}) \xi / (\Gamma + \Gamma_{12})\}$. It is evident that, compared to the ground state relaxation rate scheme, the two-photon detuning scheme is more efficient in generating entanglement and exhibits greater sensitivity to variations in optical density, as the presence of $\Gamma_{12}$ induces significant attenuation of optical density.

Alternatively, the generation of entanglement between light fields can also be understood in terms of quantum correlation, as entanglement represents a strong manifestation of quantum correlation. The entanglement between the two vortex light fields in this system arises from the quantum fluctuation coherence ($\hat{s}_{g_1 g_2}$) between the two low-energy states, as clearly demonstrated in Eqs. (15)-(23). This quantum coherence between the two low-energy states induces a change in the quantum fluctuations between the control and signal fields with respect to the transmission distance $\xi$, as expressed in Eqs. (28) and (29). These fluctuations become a linear combination of the quantum fluctuations of each light field and the Langevin noise term, leading to the mutual coupling of the fluctuations between the control and signal fields. As a result, the two light fields establish quantum correlation (entanglement) as they propagate through the medium. In the absence of atomic coherence ($\hat{\sigma}_{g_1 g_2}$), the control and signal fields will not couple, and consequently, no quantum correlation (entanglement) will be established between the two light fields.

To derive an analytical expression for entanglement, we assume a stable propagation of the light field in the medium, neglecting any variations. This approximation holds when the optical density $\alpha$ is sufficiently high, as the light field's attenuation occurs primarily at the medium's entrance, after which stable transmission is rapidly attained, as depicted in Fig. 2. Based on Eqs. (10) and (11), the stably transmitted light fields can be expressed as

$$\Omega_c \approx \frac{\Omega}{W} \beta_1 |c_1|^2, \tag{30}$$

$$\Omega_s \approx -\frac{\Omega}{W} \beta_1 c_1^* c_2. \tag{31}$$

Considering only the coefficients associated with self-phase interaction and the two-mode squeezing process, while neglecting less influential terms, within the two-photon detuning regime ($\Gamma_{12} = 0$), we obtain

$$P_1 \approx -\frac{2i\alpha\Gamma^2\delta^2}{(-i+\delta)^3 \Omega^4}, \quad S_1 \approx \frac{2i\alpha\Gamma^2\delta^2}{(-i+\delta)^3 \Omega^4},$$

$$R_2 \approx \frac{2i\alpha\Gamma^2\delta^2(i+\delta)}{(-i+\delta)^4 \Omega^4}, \quad Q_2 \approx \frac{2\alpha\Gamma^2\delta^2(1-i\delta)}{(-i+\delta)^4 \Omega^4}. \tag{32}$$

By evaluating the entanglement expression (A32), an approximate analytical formula for the entanglement degree $V$ can be derived as

$$V = 4 - \frac{4e^{-\lambda}(1+\delta^2)^4[\cos(\delta\lambda) - \cosh(\lambda)]}{\eta^2} - $$
$$4e^{-2\lambda}\sqrt{e^{2\lambda}(16\zeta^2+\eta^2)(-\cos(\delta\lambda)+\cosh\lambda)\left[(-16\zeta^2+\eta^2)\cos(\delta\lambda)+(16\zeta^2+\eta^2)\cosh\lambda + 8\zeta\eta\sin(\delta\lambda)\right]}, \tag{33}$$

where $\lambda = \frac{4\alpha\delta^2\Gamma^2(1-6\delta^2+\delta^4)}{(1+\delta^2)^4 \Omega^4}$, $\zeta = \delta(-1+\delta^2)$, $\eta = (1-6\delta^2+\delta^4)$. This formula qualitatively captures the variation in the entanglement $V$, with the third term of Eq. (33) decreasing monotonically with $\delta$, thereby enhancing entanglement, while the second term increases monotonically with $\delta$, leading to entanglement degradation. Consequently, there exists an optimal value of $\delta$ that minimizes $V$. The influence of other parameters on entanglement can similarly be derived from Eq. (33).

The coherent ground state superposition proposed in this study can be experimentally realized through the fractional or partial stimulated Raman adiabatic passage (STIRAP) [54]. It has been demonstrated that coherent ground state superpositions can be robustly and controllably generated in a four-state tripod system through a sequence of three laser pulses [55,56]. This technique relies on the existence and interaction of two degenerate dark states. By adjusting the relative delay of the pulses, the mixing of dark states can be controlled, thereby enabling the creation of arbitrary superposition states. Moreover, this three-level $\Lambda$ scheme can be experimentally implemented. For example, in the $^{87}Rb$ atom, the excited state $|e\rangle$ can correspond to the $|5P_{1/2}, F=1, m_F=0\rangle$ state. The two ground states $|g_1\rangle$ and $|g_2\rangle$ can correspond to $|5S_{1/2}, F=1, m_F=1\rangle$ and $|5S_{1/2}, F=1, m_F=-1\rangle$, respectively [57].

## 5. Conclusion

This paper presents a scheme for generating two entangled vortex beams from a single vortex beam within a coherently prepared three-level $\Lambda$-type medium, utilizing the Raman scattering process. This approach is distinct from conventional SPDC techniques. We systematically analyzed the entanglement characteristics of the system under conditions where two-photon detuning and ground state relaxation rate exist independently, as well as when these both parameters are simultaneously present. Our numerical results reveal that two-photon detuning yields a higher degree of output entanglement compared to ground state relaxation rate, and we identified the optimal parameter values for maximizing entanglement. Additionally, we investigated the influence of input field Rabi frequency and optical density on the degree of entanglement. We demonstrate that this entanglement arises from the quantum fluctuation correlation between the two light fields, induced by atomic coherence. The approach presented in this study for realizing vortex light entanglement in coherently prepared media can be expanded to more intricate atomic systems, enabling the entanglement of multiple vortex beams. In contrast to DV systems, CV systems enable the realization of large-scale entanglement. The application of CV entanglement of vortex light in fields such as quantum teleportation, quantum dense coding, and quantum computing is expected to demonstrate its significant potential in enhancing the capacity and efficiency of quantum information processing.

## Appendix

This appendix provides a comprehensive and detailed account of the theoretical calculations performed. By solving the Heisenberg-Langevin equation (2) for the three-level $\Lambda$-type system, we derive the equation of motion for the atomic operator:

$$\frac{\partial}{\partial t}\hat{\sigma}_{eg_1} = \left(-\frac{\Gamma}{2} - i\Delta_s\right)\hat{\sigma}_{eg_1} - i\frac{\hat{\Omega}_s^\dagger}{2}\left(\hat{\sigma}_{g_1 g_1} - \hat{\sigma}_{ee}\right) - i\frac{\hat{\Omega}_c^\dagger}{2}\hat{\sigma}_{g_2 g_1} + \hat{F}_{eg_1}, \tag{A1}$$

$$\frac{\partial}{\partial t}\hat{\sigma}_{eg_2} = \left(-\frac{\Gamma + \Gamma_{12}}{2} - i\Delta_c\right)\hat{\sigma}_{eg_2} - i\frac{\hat{\Omega}_c^\dagger}{2}\left(\hat{\sigma}_{g_2 g_2} - \hat{\sigma}_{ee}\right) - i\frac{\hat{\Omega}_s^\dagger}{2}\hat{\sigma}_{g_1 g_2} + \hat{F}_{eg_2}, \tag{A2}$$

$$\frac{\partial}{\partial t}\hat{\sigma}_{g_2 g_1} = \left(-\frac{\Gamma_{12}}{2} - \gamma_\phi - i\delta\right)\hat{\sigma}_{g_2 g_1} + i\frac{\hat{\Omega}_s^\dagger}{2}\hat{\sigma}_{g_2 e} - i\frac{\hat{\Omega}_c}{2}\hat{\sigma}_{eg_1} + \hat{F}_{g_2 g_1}, \tag{A3}$$

$$\frac{\partial}{\partial t}\hat{\sigma}_{g_1 g_1} = \Gamma_1 \hat{\sigma}_{ee} + \Gamma_{12}\hat{\sigma}_{g_2 g_2} - i\frac{\hat{\Omega}_s}{2}\hat{\sigma}_{eg_1} + i\frac{\hat{\Omega}_s^\dagger}{2}\hat{\sigma}_{g_1 e} + \hat{F}_{g_1 g_1}, \tag{A4}$$

$$\frac{\partial}{\partial t}\hat{\sigma}_{g_2 g_2} = \Gamma_2 \hat{\sigma}_{ee} - \Gamma_{12}\hat{\sigma}_{g_2 g_2} - i\frac{\hat{\Omega}_c}{2}\hat{\sigma}_{eg_2} + i\frac{\hat{\Omega}_c^\dagger}{2}\hat{\sigma}_{g_2 e} + \hat{F}_{g_2 g_2}, \tag{A5}$$

$$\frac{\partial}{\partial t}\hat{\sigma}_{ee} = -\Gamma\hat{\sigma}_{ee} + i\frac{\hat{\Omega}_s}{2}\hat{\sigma}_{eg_1} + i\frac{\hat{\Omega}_c}{2}\hat{\sigma}_{eg_2} - i\frac{\hat{\Omega}_s^\dagger}{2}\hat{\sigma}_{g_1 e} - i\frac{\hat{\Omega}_c^\dagger}{2}\hat{\sigma}_{g_2 e} + \hat{F}_{ee}, \tag{A6}$$

$$\frac{\partial}{\partial t}\hat{\sigma}_{g_1 g_2} = \frac{\partial}{\partial t}\hat{\sigma}_{g_2 g_1}^\dagger, \tag{A7}$$

$$\frac{\partial}{\partial t}\hat{\sigma}_{g_2 e} = \frac{\partial}{\partial t}\hat{\sigma}_{eg_2}^\dagger, \tag{A8}$$

$$\frac{\partial}{\partial t}\hat{\sigma}_{g_1 e} = \frac{\partial}{\partial t}\hat{\sigma}_{eg_1}^\dagger. \tag{A9}$$

By substituting the atomic operator equations of motion (A1)-(A9) into the Maxwell-Schrödinger equations (3) and (4), we derive a set of coupling equations that describe the interaction between atomic and field operators. Employing the mean-field approximation $\hat{O} = O + \hat{o}$, and disregarding the higher-order terms $\hat{o}_\mu \hat{o}_\nu$ $(\mu, \nu \in c, s, e, g_1, g_2)$ of the quantum fluctuation operators, the operator can be decomposed into its mean-field and quantum fluctuation components. Solving these two components separately allows us to ultimately determine the quantum entanglement of the output light fields.

### 1. Mean-field solution

The two ground states of the medium are prepared into a coherent superposition state:

$$|\psi\rangle = c_1 |g_1\rangle + c_2 |g_2\rangle. \tag{A10}$$

Considering the weak interaction between fields and atoms. The light field can then be treated as a perturbation, and by neglecting higher-order terms, we obtain $\sigma_{ee} \approx 0$, $\sigma_{g_1 g_1} \approx |c_1|^2$, $\sigma_{g_2 g_2} \approx |c_2|^2$, $\sigma_{g_1 g_2} \approx c_1^* c_2$. However, previous studies have demonstrated that this approximation can be relaxed when the light field is transmitted stably in the medium, i.e., when it remains invariant with respect to $\xi$ [46]. Our focus is solely on the steady-state solution, which does not vary with time. Thus, the time-dependent partial derivative terms in the equations can be neglected. Substituting these results into the mean-field equations of motion (5) and (6), we obtain a set of fully decoupled equations, which provide the steady-state solutions for the coherences $\sigma_{g_1 e}$ and $\sigma_{g_2 e}$ as follows

$$\sigma_{g_1 e} = -\frac{|c_1|^2 \Omega_s + c_1^* c_2 \Omega_c}{i\Gamma + 2\Delta_s}, \tag{A11}$$

$$\sigma_{g_2 e} = -\frac{c_1 c_2^* \Omega_s + |c_1|^2 \Omega_c}{i(\Gamma + \Gamma_{12}) + 2\Delta_c}. \tag{A12}$$

Similarly, the steady-state equation of the mean-field component of the Maxwell-Schrödinger equation is described as follows

$$\frac{\partial}{\partial z}\Omega_s = i\frac{\Gamma \alpha}{2L}\sigma_{g_1 e}, \tag{A13}$$

$$\frac{\partial}{\partial z}\Omega_c = i\frac{\Gamma \alpha}{2L}\sigma_{g_2 e}, \tag{A14}$$

substituting Eqs. (A11) and (A12) into Eqs. (A13) and (A14), we derive the coupling equations governing the propagation of light fields in the medium, as follows

$$\frac{\partial \Omega_s}{\partial z} = -i\beta_1 \left( |c_1|^2 \Omega_s + c_1^* c_2 \Omega_c \right), \tag{A15}$$

$$\frac{\partial \Omega_c}{\partial z} = -i\beta_2 \left( c_1 c_2^* \Omega_s + |c_2|^2 \Omega_c \right), \tag{A16}$$

where $\beta_1 = \frac{\Gamma \alpha}{2L(2\Delta_s + i\Gamma)}$ and $\beta_2 = \frac{\Gamma \alpha}{2L\left[2\Delta_c + i(\Gamma + \Gamma_{12})\right]}$.

By applying the initial boundary conditions, $\Omega_c(0) = \Omega$, $\Omega_s(0) = 0$, and solving Eqs. (A15) and (A16), we derive the Eqs. (10) and (11) for the light field propagation in the medium.

## 2. Quantum fluctuations solution

As our study focuses exclusively on the steady-state process, the time derivative term is set to zero. Consequently, Eqs. (15)–(23) can be expressed in matrix form $\mathbf{M}_1 \mathbf{y} + \mathbf{M}_2 \mathbf{a} + \mathbf{r} = \mathbf{0}$, where $\mathbf{y} = (\hat{s}_{eg_1}, \hat{s}_{eg_2}, \hat{s}_{g_2 g_1}, \hat{s}_{g_1 g_1}, \hat{s}_{g_2 g_2}, \hat{s}_{ee}, \hat{s}_{g_1 g_2}, \hat{s}_{g_2 e}, \hat{s}_{g_1 e})^\mathrm{T}$ represents atomic operator fluctuations, $\mathbf{a} = (\hat{a}_s, \hat{a}_s^\dagger, \hat{a}_c, \hat{a}_c^\dagger)^\mathrm{T}$ denotes light field operator fluctuations, and $\mathbf{r} = (\hat{F}_{eg_1}, \hat{F}_{eg_2}, \hat{F}_{g_2 g_1}, \hat{F}_{g_1 g_1}, \hat{F}_{g_2 g_2}, \hat{F}_{ee}, \hat{F}_{g_1 g_2}, \hat{F}_{g_2 e}, \hat{F}_{g_1 e})^\mathrm{T}$ corresponds to the Langevin noise operator. The matrices $\mathbf{M}_1$ and $\mathbf{M}_2$ are expressed as follows

$$\mathbf{M}_1 = \begin{pmatrix}
-\tilde{\gamma}_{13}^* & 0 & -i\frac{\Omega_c^*}{2} & -i\frac{\Omega_s^*}{2} & 0 & i\frac{\Omega_s^*}{2} & 0 & 0 & 0 \\
0 & -\tilde{\gamma}_{23}^* & 0 & 0 & -i\frac{\Omega_c^*}{2} & i\frac{\Omega_c^*}{2} & -i\frac{\Omega_s^*}{2} & 0 & 0 \\
-i\frac{\Omega_c}{2} & 0 & -\tilde{\gamma}_{12}^* & 0 & 0 & 0 & 0 & i\frac{\Omega_s^*}{2} & 0 \\
-i\frac{\Omega_s}{2} & 0 & 0 & 0 & \Gamma_{12} & \frac{\Gamma}{2} & 0 & 0 & i\frac{\Omega_s^*}{2} \\
0 & -i\frac{\Omega_c}{2} & 0 & 0 & -\Gamma_{12} & \frac{\Gamma}{2} & 0 & i\frac{\Omega_c^*}{2} & 0 \\
0 & 0 & 0 & 1 & 1 & 1 & 0 & 0 & 0 \\
0 & -i\frac{\Omega_s}{2} & 0 & 0 & 0 & 0 & -\tilde{\gamma}_{12} & 0 & i\frac{\Omega_c^*}{2} \\
0 & 0 & i\frac{\Omega_s}{2} & 0 & i\frac{\Omega_c}{2} & -i\frac{\Omega_c}{2} & 0 & -\tilde{\gamma}_{23} & 0 \\
0 & 0 & 0 & i\frac{\Omega_s}{2} & 0 & -i\frac{\Omega_s}{2} & i\frac{\Omega_c}{2} & 0 & -\tilde{\gamma}_{13}
\end{pmatrix}_{9\times 9}, \tag{A17}$$

where $\tilde{\gamma}_{13} = \Gamma/2 - i\Delta_s$, $\tilde{\gamma}_{23} = (\Gamma + \Gamma_{12})/2 - i\Delta_c$ and $\tilde{\gamma}_{12} = (\Gamma_{12}/2 + \gamma_\phi) - i\Delta_c$. Considering particle number conservation, Eq. (20) is replaced by $\sigma_{11} + \sigma_{22} + \sigma_{33} = 1$.

$$\mathbf{M}_2 = \frac{g}{2}\begin{pmatrix} 0 & -i|c_1|^2 & 0 & -ic_1c_2^* \\ 0 & -ic_1^*c_2 & 0 & -i|c_2|^2 \\ 0 & i\sigma_{g_2e} & -i\sigma_{eg_1} & 0 \\ -i\sigma_{eg_1} & i\sigma_{g_1e} & 0 & 0 \\ 0 & 0 & -i\sigma_{eg_2} & i\sigma_{g_2e} \\ 0 & 0 & 0 & 0 \\ -i\sigma_{eg_2} & 0 & 0 & i\sigma_{g_1e} \\ ic_1c_2^* & 0 & i|c_2|^2 & 0 \\ i|c_1|^2 & 0 & ic_1^*c_2 & 0 \end{pmatrix}_{9\times 4}.\quad (A18)$$

Solving for $\mathbf{y} = -\mathbf{M}_1^{-1}(\mathbf{M}_2\mathbf{a}+\mathbf{r})$ allows us to express the atomic fluctuation operator in terms of the field operator.

$$\hat{s}_{g_1e} = \mathbf{y}(9) = A_1\hat{a}_s + B_1\hat{a}_s^\dagger + C_1\hat{a}_c + D_1\hat{a}_c^\dagger + \hat{f}_{g_1e}, \quad (A19)$$

$$\hat{s}_{g_2e} = \mathbf{y}(8) = A_2\hat{a}_s + B_2\hat{a}_s^\dagger + C_2\hat{a}_c + D_2\hat{a}_c^\dagger + \hat{f}_{g_2e}, \quad (A20)$$

where $\hat{f}_{g_1e}$ and $\hat{f}_{g_2e}$ denote effective Langevin noise operators, derived from the following equations:

$$\hat{f}_{g_1e} = (-\mathbf{M}_1^{-1}\mathbf{r})_9, \quad (A21)$$

$$\hat{f}_{g_2e} = (-\mathbf{M}_1^{-1}\mathbf{r})_8. \quad (A22)$$

Conversely, the quantum fluctuation component of the Maxwell-Schrödinger equation for steady-state transport can be derived as

$$\frac{\partial}{\partial \xi}\hat{a}_s = i\frac{\Gamma\alpha}{2g}\hat{s}_{g_1e}, \quad (A23)$$

$$\frac{\partial}{\partial \xi}\hat{a}_c = i\frac{\Gamma\alpha}{2g}\hat{s}_{g_2e}, \quad (A24)$$

where $\xi = z/L$, which is a dimensionless length.

Substituting Eqs. (A19) and (A20) into Eqs. (A23) and (A24) yields a compact form where the field fluctuation operator satisfies the following equations:

$$\frac{\partial}{\partial \xi}\mathbf{a} = \mathbf{C}\mathbf{a} + \mathbf{N}. \quad (A25)$$

The matrices $\mathbf{C}$ and $\mathbf{N}$ are defined as follows

$$\mathbf{C} = i\frac{\Gamma\alpha}{2g}\begin{pmatrix} A_1 & B_1 & C_1 & D_1 \\ -B_1^* & -A_1^* & -D_1^* & -C_1^* \\ A_2 & B_2 & C_2 & D_2 \\ -B_2^* & -A_2^* & -D_2^* & -C_2^* \end{pmatrix} \equiv \begin{pmatrix} P_1 & Q_1 & R_1 & S_1 \\ Q_1^* & P_1^* & S_1^* & R_1^* \\ P_2 & Q_2 & R_2 & S_2 \\ Q_2^* & P_2^* & S_2^* & R_2^* \end{pmatrix}, \quad (A26)$$

$$\mathbf{N} = i\frac{\Gamma\alpha}{2g}(\hat{f}_{g_1e}, -\hat{f}_{g_1e}^\dagger, \hat{f}_{g_2e}, -\hat{f}_{g_2e}^\dagger)^{\mathrm{T}}. \quad (A27)$$

### 3. Quantum correlation solution

To calculate the system's entanglement properties, it is essential to determine the field correlations. From Eqs. (A25)–(A27), the equations of motion governing all field correlations can be derived as follows

$$\frac{\partial}{\partial \xi}\langle \mathbf{a}\mathbf{a}^\dagger \rangle = \mathbf{C}\langle \mathbf{a}\mathbf{a}^\dagger \rangle + \langle \mathbf{a}\mathbf{a}^\dagger \rangle \mathbf{C}^\dagger + \mathbf{Z}, \quad (A28)$$

where the matrix $\mathbf{Z}$ represents the correlation of the Langevin noise operator and is expressed as

$$\mathbf{Z} = \langle \mathbf{NN}^\dagger \rangle = \frac{\Gamma\alpha}{4}\left(\mathbf{V}\mathcal{D}\mathbf{V}^\dagger\right), \tag{A29}$$

where we define $\alpha = g^2 NL/(c\Gamma)$. The correlation between any two Langevin noise operators is described by the equation $\langle F_\mu F_\nu \rangle = \mathcal{D}_{\mu\nu} c/(NL)$, where $\mathcal{D}_{\mu\nu}$ is the diffusion coefficient, determined by Einstein's generalized relation [58-60]. The matrix $\mathbf{V}$ takes the form:

$$\mathbf{V} \equiv \begin{pmatrix} T_{91} & T_{92} & T_{93} & T_{94} & T_{95} & T_{96} & T_{97} & T_{98} & T_{99} \\ -T_{11} & -T_{12} & -T_{13} & -T_{14} & -T_{15} & -T_{16} & -T_{17} & -T_{18} & -T_{19} \\ T_{81} & T_{82} & T_{83} & T_{84} & T_{85} & T_{86} & T_{87} & T_{88} & T_{89} \\ -T_{21} & -T_{22} & -T_{23} & -T_{24} & -T_{25} & -T_{26} & -T_{27} & -T_{28} & -T_{29} \end{pmatrix}_{4\times 9}, \tag{A30}$$

where $\mathbf{T} \equiv -\mathbf{M}_1^{-1}$. The corresponding diffusion coefficient $\mathcal{D}$ is expressed in matrix form as

$$\mathcal{D} = \begin{pmatrix} 0 & 0 & \gamma_\phi \sigma_{eg_2} & 0 & 0 & 0 & 0 & 0 & 0 \\ 0 & \Gamma_{12}\sigma_{ee} & 0 & -\Gamma_{12}\sigma_{eg_2} & \Gamma_{12}\sigma_{eg_2} & 0 & (\Gamma_{12}+\gamma_\phi)\sigma_{eg_1} & 0 & 0 \\ \gamma_\phi \sigma_{g_2 e} & 0 & \Gamma_2 \sigma_{ee} + 2\gamma_\phi \sigma_{g_2 g_2} & 0 & 0 & 0 & 0 & 0 & 0 \\ 0 & -\Gamma_{12}\sigma_{g_2 e} & 0 & \Gamma_1\sigma_{ee}+\Gamma_{12}\sigma_{g_2 g_2} & -\Gamma_{12}\sigma_{g_2 g_2} & 0 & -\Gamma_{12}\sigma_{g_2 g_1} & -\Gamma_1\sigma_{eg_2} & -\Gamma_1\sigma_{eg_1} \\ 0 & \Gamma_{12}\sigma_{g_2 e} & 0 & -\Gamma_{12}\sigma_{g_2 g_2} & \Gamma_2\sigma_{ee}+\Gamma_{12}\sigma_{g_2 g_2} & 0 & \Gamma_{12}\sigma_{g_2 g_1} & -\Gamma_2\sigma_{eg_2} & -\Gamma_2\sigma_{eg_1} \\ 0 & 0 & 0 & 0 & 0 & 0 & 0 & 0 & 0 \\ 0 & (\Gamma_{12}+\gamma_\phi)\sigma_{g_1 e} & 0 & -\Gamma_{12}\sigma_{g_1 g_2} & \Gamma_{12}\sigma_{g_1 g_2} & 0 & \Gamma_1\sigma_{ee}+\Gamma_{12}\sigma_{g_2 g_2}+(\Gamma_{12}+2\gamma_\phi)\sigma_{g_1 g_1} & 0 & 0 \\ 0 & 0 & 0 & -\Gamma_1\sigma_{g_2 e} & -\Gamma_2\sigma_{g_2 e} & 0 & 0 & \Gamma_2\sigma_{ee}+\Gamma\sigma_{g_2 g_2} & (\Gamma-\gamma_\phi)\sigma_{g_2 g_1} \\ 0 & 0 & 0 & -\Gamma_1\sigma_{g_1 e} & -\Gamma_2\sigma_{g_1 e} & 0 & 0 & (\Gamma-\gamma_\phi)\sigma_{g_1 g_2} & \Gamma_1\sigma_{ee}+\Gamma_{12}\sigma_{g_2 g_2}+\Gamma\sigma_{g_1 g_1} \end{pmatrix}_{9\times 9}$$

(A31)

Consequently, solving Eq. (A28) provides the explicit form of the entanglement expression (27) as

$$V = 4\left(1 + \langle \mathbf{aa}^\dagger \rangle_{22} + \langle \mathbf{aa}^\dagger \rangle_{44} - 2\left|\langle \mathbf{aa}^\dagger \rangle_{14}\right|\right). \tag{A32}$$

**Disclosures**

The authors declare no conflicts of interest.

**Data availability**

No additional data are available.

**Acknowledgements**

The authors acknowledge the National Natural Science Foundation of China (Grant Nos. 12074027, 12474353).

**References**

[1] A. Einstein, B. Podolsky, N. Rosen, "Can quantum-mechanical description of physical reality be considered complete?" Phys. Rev. 47 (10), 777-780 (1935).
[2] A. Aspect, J. Dalibard, G. Roger, "Experimental test of Bell's inequalities using time-varying analyzers," Phys. Rev. Lett. 49 (25), 1804 (1982).
[3] M. H. Rubin, D. N. Klyshko, Y. H. Shih, and A. V. Sergienko, "Theory of two-photon entanglement in type-II optical parametric down-conversion," Phys. Rev. A 50, 5122 (1994).
[4] L. Mandel, "Quantum effects in one-photon and two-photon interference," Rev. Mod. Phys. 71, S274 (1999).
[5] P. Kolchin, S. Du, C. Belthangady, G. Y. Yin, and S. E. Harris, "Generation of Narrow-

Bandwidth Paired Photons: Use of a Single Driving Laser," Phys. Rev. Lett. 97, 113602 (2006).
[6] J. M. Wen and M. H. Rubin, "Transverse effects in paired-photon generation via an electromagnetically induced transparency medium. I. Perturbation theory," Phys. Rev. A 74, 023808 (2006).
[7] J. Wen, S. W. Du, and M. H. Rubin, "Biphoton generation in a two-level atomic ensemble," Phys. Rev. A 75, 033809 (2007).
[8] H. S. Zhong, Y. Li, W. Li, L. C. Peng, Z. E. Su, Y. Hu, Y. M. He, X. Ding, W. J. Zhang, H. Li, L. Zhang, Z. Wang, L. X. You, X. L. Wang, X. Jiang, L. Li, Y. A. Chen, N. L. Liu, C. Y. Lu, and J. W. Pan, "12-photon entanglement and scalable scattershot boson sampling with optimal entangled-photon pairs from parametric down-conversion," Phys. Rev. Lett. 121(25), 250505 (2018).
[9] L.-A. Wu, M. Xiao, H. J. Kimble, "Squeezed states of light from an optical parametric oscillator," J. Opt. Soc. Am. B 4 (10), 1465-1475 (1987).
[10] R. E. Slusher, L. W. Hollberg, B. Yurke, J. C. Mertz, and J. F. Valley, "Observation of squeezed states generated by four-wave mixing in an optical cavity," Phys. Rev. Lett. 55, 2409 (1985).
[11] L.-A. Wu, H. J. Kimble, J. L. Hall, and H. Wu, "Generation of squeezed states by parametric down conversion," Phys. Rev. Lett. 57, 2520 (1986).
[12] U. L. Andersen, T. Gehring, C. Marquardt, and G. Leuchs, "30 years of squeezed light generation," Phys. Scr. 91, 053001 (2016).
[13] A. Furusawa, J. L. Sørensen, S. L. Braunstein, C. A. Fuchs, H. J. Kimble, and E. S. Polzik, "Unconditional quantum teleportation," Science 282, 706 (1998).
[14] Y. L. Chuang, R. K. Lee, and I. A. Yu, "Optical-density-enhanced squeezed-light generation without optical cavities," Phys. Rev. A 96, 053818 (2017).
[15] Y. L. Chuang, R. K. Lee, and I. A. Yu, "Generation of quantum entanglement based on electromagnetically induced transparency media," Opt. Express 29(3), 3928 (2021).
[16] K. Zhang, W. Wang, S. S. Liu, X. Z. Pan, J. J. Du, Y. B. Lou, S. Yu, S. C. Lv, N. Treps, C. Fabre, and J. T. Jing, "Reconfigurable hexapartite entanglement by spatially multiplexed four-wave mixing processes," Phys. Rev. Lett. 124 (9), 090501 (2020).
[17] M. Chen, N. C. Menicucci, O. Pfister, "Experimental realization of multipartite entanglement of 60 modes of a quantum optical frequency comb," Phys. Rev. Lett. 112 (12), 120505 (2014).
[18] X. L. Su, Y. P Zhao, S. H. Hao, X. J. Jia, C. D. Xie, and K. C. Peng, "Experimental preparation of eight-partite cluster state for photonic qumodes," Opt. Lett. 37 (24), 5178-5180 (2012).
[19] Q. Glorieux, R. Dubessy, S. Guibal, L. Guidoni, J.-P. Likforman, and T. Coudreau, "Double-Λ microscopic model for entangled light generation by four-wave mixing," Phys. Rev. A 82, 033819 (2010).
[20] J. W. Pan, M. Daniell, S. Gasparoni, G. Weihs, and A. Zeilinger, "Experimental Demonstration of Four-Photon Entanglement and High-Fidelity Teleportation," Phys. Rev. Lett. 86(20), 4435 (2001).
[21] N. A. Ron, M. Carmi, and R. Bekenstein, "Atom-Atom entanglement generation via collective states of atomic rings," Phys. Rev. Res. 6, L042051 (2024).
[22] C. Monroe, D. M. Meekhof, B. E. King, D. J. Wineland, "A 'Schrödinger cat' superposition state of an atom," Science 272 (5265), 1131-1136 (1996).
[23] M. Erhard, R. Fickler, M. Krenn, and A. Zeilinger, "Twisted photons: new quantum perspectives in high dimensions," Light: Sci. Appl. 7, 17146 (2018).
[24] J. Wang, J. Y. Yang, I. M. Fazal, N. Ahmed, Y. Yan, H. Huang, Y. X. Ren, Y. Yue, S. Dolinar, M. Tur, and A. E. Willner, "Terabit free-space data transmission employing orbital angular momentum multiplexing," Nat. Photonics 6(7), 488-496 (2012).
[25] G. D. Xie, L. Li, Y. X. Ren, H. Huang, Y. Yan, N. Ahmed, Z. Zhao, M. P. J. Lavery, N. Ashrafi, S. Ashrafi, R. Bock, M. Tur, A. F. Molisch, and A. E. Willner, "Performance metrics and design considerations for a free-space optical orbital-angular-momentum–multiplexed communication link," Optica 2(4), 357-365 (2015).

[26] Y. Yan, G. D. Xie, M. P. J. Lavery, H. Huang, N. Ahmed, C. J. Bao, Y. X. Ren, Y. W. Cao, L. Li, Z. Zhao, A. F. Molisch, M. Tur, M. J. Padgett, and A. E. Willner, "High-capacity millimetre-wave communications with orbital angular momentum multiplexing," Nat. Commun. 5(1), 4876 (2014).

[27] A. Mair, A. Vaziri, G. Weihs, A. Zeilinger, "Entanglement of the orbital angular momentum states of photons," Nature 412(6844), 313-316 (2001).

[28] Q. F. Chen, B. S. Shi, Y. S. Zhang, and G. C. Guo, "Entanglement of the orbital angular momentum states of the photon pairs generated in a hot atomic ensemble," Phys. Rev. A 78, 053810 (2008).

[29] Z. Q. Zhou, Y. L. Hua, X. Liu, G. Chen, J. S. Xu, Y. J. Han, C. F. Li, and G. C. Guo, "Quantum storage of three-dimensional orbital-angular-momentum entanglement in a crystal," Phys. Rev. Lett. 115, 070502 (2015).

[30] B. C. Hiesmayr, M. J. A. de Dood, and W. Löffler, "Observation of four-photon orbital angular momentum entanglement," Phys. Rev. Lett. 116, 073601 (2016).

[31] W. Zhang, D. S. Ding, M. X. Dong, S. Shi, K. Wang, S. L. Liu, Y. Li, Z.Y. Zhou, B. S. Shi, and G. Guo, "Experimental realization of entanglement in multiple degrees of freedom between two quantum memories," Nat. Commun. 7, 13514 (2016).

[32] M. Lassen, G. Leuchs, and U.L. Andersen, "Continuous variable entanglement and squeezing of orbital angular momentum states," Phys. Rev. Lett. 102(16), 163602 (2009).

[33] X. Z. Pan, S. Yu, Y. F. Zhou, K. Zhang, K. Zhang, S. C. Lv, S. J. Li, W. Wang, and J. T. Jing, "Orbital-angular-momentum multiplexed continuous-variable entanglement from four-wave mixing in hot atomic vapor," Phys. Rev. Lett. 123, 070506 (2019).

[34] S. J. Li, X. Z. Pan, Y. Ren, H. Z. Liu, S. Yu, and J. T. Jing, "Deterministic generation of orbital-angular-momentum multiplexed tripartite entanglement," Phys. Rev. Lett. 124, 083605 (2020).

[35] A. C. Dada, J. Leach, G.S. Buller, M. J. Padgett, and E. Andersson, "Experimental high-dimensional two-photon entanglement and violations of generalized Bell inequalities," Nat. Phys. 7, 677 (2011).

[36] M. Mafu, A. Dudley, S. Goyal, D. Giovannini, M. McLaren, M. J. Padgett, T. Konrad, F. Petruccione, N. Lütkenhaus, and A. Forbes, "Higher-dimensional orbital-angular-momentum-based quantum key distribution with mutually unbiased bases," Phys. Rev. A 88, 032305 (2013).

[37] A. M. Marino, V. Boyer, R. C. Pooser, P. D. Lett, K. Lemons, and K. M. Jones, "Delocalized correlations in twin light beams with orbital angular momentum," Phys. Rev. Lett. 101, 093602 (2008).

[38] D.S. Ding, W. Zhang, Z. Y. Zhou, S. Shi, G. Y. Xiang, X. S. Wang, Y. K. Jiang, B. S. Shi, and G. C. Guo, "Quantum Storage of Orbital Angular Momentum Entanglement in an Atomic Ensemble," Phys. Rev. Lett. 114, 050502 (2015).

[39] R. W. Boyd, "Nonlinear optics," 4th ed. (Academic Press, London, 2020).

[40] R. Hazra, M.M. Hossain, "Study of atomic populations, electromagnetically induced transparency, and dispersive signals in a Λ-type system under various decoherence effects," Ukr. J. Phys. 64(3), 197 (2019).

[41] P. Kolchin, "Electromagnetically-induced-transparency-based paired photon generation," Phys. Rev. A 75, 033814 (2007).

[42] M. O. Scully, M. S. Zubairy, "Quantum Optics," (Cambridge University Press, Cambridge, 1997).

[43] E. Fiorelli, M. Müller, I. Lesanovsky and Federico Carollo, "Mean-field dynamics of open quantum systems with collective operator-valued rates: validity and application," New J. Phys. 25, 083010 (2023).

[44] P. Kirton and J. Keeling, "Suppressing and Restoring the Dicke Superradiance Transition by Dephasing and Decay," Phys. Rev. Lett. 118, 123602 (2017).

[45] P. Wang 1, and R. Fazio, "Dissipative phase transitions in the fully connected Ising model with p-spin interaction," Phys. Rev. A 103, 013306 (2021).

[46] E. Paspalakis, N. J. Kylstra, and P. L. Knight, "Propagation and nonlinear generation dynamics in a coherently prepared four-level system," Phys. Rev. A 65, 053808 (2002).

[47] M. Fleischhauer, A. Imamoglu, and J. P. Marangos, "Electromagnetically induced transparency: Optics in coherent media," Rev. Mod. Phys. 77(2), 633-673 (2005).
[48] L. M. Duan, G. Giedke, J. I. Cirac, and P. Zoller, "Inseparability criterion for continuous variable systems," Phys. Rev. Lett. 84(12), 2722–2725 (2000).
[49] R. Simon, "Peres-horodecki separability criterion for continuous variable systems," Phys. Rev. Lett. 84(12), 2726–2729 (2000).
[50] S. L. Braunstein and P. van Loock, "Quantum information with continuous variables," Rev. Mod. Phys. 77, 513 (2005).
[51] S.S. Liu, Y. B. Lou, and J.T. Jing, "Orbital angular momentum multiplexed deterministic all-optical quantum teleportation," Nat. Commun. 11, 3875 (2020).
[52] B. M. Sparkers, J. Bernu, M. Hosseini, J. Geng, Q. Glorieux, P. A. Altin, P. K. Lam, N. P. Robins, and B. C. Buchler, "Gradient echo memory in an ultra-high optical depth cold atomic ensemble," New J. Phys. 15, 085027 (2013).
[53] F. Blatt, T. Halfmann, and T. Peters, "One-dimensional ultracold medium of extreme optical depth," Opt. Lett. 39, 446 (2014).
[54] N. V. Vitanov, A. A. Rangelov, B. W. Shore, and K. Bergmann, "Stimulated Raman adiabatic passage in physics, chemistry, and beyond," Rev. Mod. Phys. 89, 015006 (2017).
[55] R. G. Unanyan, M. Fleischhauer, B. W. Shore, and K. Bergmann, "Robust creation and phase-sensitive probing of superposition states via stimulated Raman adiabatic passage STIRAP with degenerate dark states," Opt. Commun. 155, 144 (1998).
[56] R. G. Unanyan, B. W. Shore, and K. Bergmann, "Laser-driven population transfer in four-level atoms: Consequences of non-Abelian geometrical adiabatic phase factors," Phys. Rev. A 59, 2910 (1999).
[57] L. G. Si, W. X. Yang, and X. X. Yang, "Ultraslow temporal vector optical solitons in a cold four-level tripod atomic system," J. Opt. Soc. Am. B 26, 478 (2009).
[58] P. Barberis-Blostein and N. Zagury, "Field correlations in electromagnetically induced transparency," Phys. Rev. A 70, 053827 (2004).
[59] L. Davidovich, "Sub-Poissonian processes in quantum optics," Rev. Mod. Phys. 68, 127 (1996).
[60] C. Cohen-Tannoudji, J. Dupont-Roc, G. Grynberg, "Atom-Photon Interactions: Basic Processes and Applications," (Wiley, New York, 1998).